# Influence of a non-uniform current density profile on quasiparticle excitations and thermally activated flux creep in YBa$_2$Cu$_3$O$_{7-\delta}$

S.A. Sergeenkov[1,*], V.V. Gridin[2], M. Ausloos[1]

[1] SUPRAS, Institute of Physics, B5, Sart Tilman, University of Liège, B-4000 Liège, Belgium (e-mail: ausloos@gw.unipc.ulg.ac.be)
[2] Department of Physics, University of the Witwatersrand, Johannesburg, South Africa

**Abstract.** The transport current density, flowing radially from the center of a superconducting disk to its perimeter, in a so-called Corbino geometry, results in a double action on the vortex motion when the applied magnetic field is perpendicular to the disk's plane. First, the depinned vortices are set into a nearly circular motion in the plane of the disk. Second, the non-uniform current density profile activates the intrinsic weak links, resulting in a non negligible proximity dominated quasiparticle contribution. In turn, these intrinsic junctions impede the circular motion of vortices giving rise to a proximity influenced thermally activated flux creep. This provides a simple technique to deconvolute the flux motion and the quasiparticle induced components of the total dissipative resistivity broadening below $T_c$. The case for a YBaCuO disk is hereby examined and measured in this context. It is shown that such a deconvolution can be made and theoretical laws for the field dependences are also obtained and confirm the data. Furthermore, the paraconductivity region just above $T_c(B)$ appears to be dominated by the macroscopic fluctuations accompanying the vortex core motion.

**PACS:** 74.50; 74.70V

## I. Introduction

It is well established [1–3] that transport properties of type II superconductors in their mixed state are determined by the two main dissipation mechanisms, quasiparticle excitations and motion of vortices (due to their normal cores). The latter process was found to contribute substantially to the transport properties of conventional (low-$T_c$) superconductors and to be of the same importance as the quansiparticle contribution. The situation in high-$T_c$ superconductor (HTS) materials, at least as far as the resistivity studies are concerned, but also for the cases of the thermoelectric power in a field [4, 5] and the Nernst effect [5, 6], is less clear. The quasiparticle contribution to dissipation, $W_q$, is expected to become especially significant in HTSs, provided that there exists [7] a nonvanishing quasiparticle density of states in the gap, $\Delta$, as for the $d$-wave symmetry gap. This, in turn, makes it difficult to study the depinning processes that give rise to the dissipation term, $W_\varphi$, due to the motion of fluxon cores [8]. The knowledge of the relative strength of $W_\varphi$ and $W_q$ is therefore essential when analyzing experimental data obtained in electrical resistivity measurements.

Subject to the short coherence length and the "clean limit nature" of HTS compounds, the fluctuation induced quasiparticle excitations also extend not only much above $T_c$, as seen from the Maki-Thompson contribution to the paraconductivity [9], but also well below $T_c$, increasing the temperature span of the resistive tail as observed [10]. Hohn et al. [11] and others [5, 6] more recently have proposed that galvano-magnetic effects studied in cuprates should be also influenced by the presence of the so-called bound (or localized) excitations inside flux cores as opposite to the unbound (extended) ones that are usually associated with the motion of the cores themselves.

The present paper addresses the deconvolution problem for the quasiparticle and vortex motion induced contributions to the potentiometric signals detected in resistivity measurement studies. In Sect. II we describe a rather simple experimental technique based on the use of a disk sample geometry, which allows to single out the $W_\phi$ and $W_q$ components from the total potential drop detected, $W = W_q + W_\phi$. This is followed by the qualitative description of how a non-uniform current density profile (dictated by the disk geometry) activates intrinsic proximity effects in a sample, proximity effects which originate probably from the local disturbances of the order parameter due to oxygen defects. A detailed analysis of the experimental data along with the theoretical interpretation is presented in Sect. III.

* *Permanent address:* Frank Laboratory of Neutron Physics, Joint Institute for Nuclear Research, 141980 Dubna, Moscow region, Russia

## II. Experimental method

It is well known [1, 2] that in the mixed state of type II superconductors, the single flux line moving with velocity, $V_\phi$, induces a potential drop, $W_{\phi 0}$, according to the phase slip mechanism [12]. For $N_\phi$ moving fluxons, the flux motion contribution to the total voltage drop detected in $W_\phi = N_\phi W_{\phi 0}$. However, there exists an additional voltage component, $W_q$, due to the presence of quasiparticle excitations even at $T < T_c$. The form of the electric field in the moving coordinate system (rotating fluxons), given by $\boldsymbol{E}(total) = \boldsymbol{E}(currents) + V_\phi \times \boldsymbol{B}$, suggests [13] to assume a linear combination of the potential drops detected for the magneto-resistance experiments in the mixed state of type II superconductors

$$W = W_q + W_\phi, \qquad (1)$$

where the $q$ and $\phi$ subscripts mark the temperature ($T$), current density ($J$), and magnetic field ($B$) dependent voltage contributions resulting from the quasiparticle and fluxon core components of the total signal ($W$), respectively. The temperature dependence of $W_q$ for $T < T_c$ is given by $W_q = \alpha_q(T\,J, B) W_n$, where $W_n$ is a normal state voltage drop extrapolated from the normal state regime (well above $T_c$) in such a way that there is no fluctuation induced contribution to $W_n = W_n(T)$ [14, 15]. We expect, on a physical ground, that $\alpha_q(T, J, B) \to 1$ for $T > T_c$. In other words, $\alpha_q$ represents, essentially, the normalization of observable dissipation below $T_c$ to that above $T_c$ when the vortex motion contribution has been accounted for. The latter is due to the vortices when they cross the line connecting the voltage contacts for $W$ [1].

Let us first demonstrate a simple experimental method which allows us to deconvolute the quasiparticle, $W_q$, and the fluxon core motion, $W_\phi$, contributions to the mixed state dissipation of a HTS. The importance of the contact arrangement on the potentiometric signals detected was considered by Clem as early as in 1970 [16]. Let us consider first a conventional, bar sample geometry (which is shown in Fig. 1). The current is introduced through electrode 1 and leaves through the lead 2. A magnetic field B is applied along the Z-direction perpendicularly to the largest planar section of the sample. In such a case, due to the constant and position independent vortex velocity $V_\phi$, the voltage $W_\phi$ developed across the 3–4 contact pair (Fig. 1) is proportional to the distance $\Delta L_{3-4}$ according to $W_\phi \propto V_\phi \Delta L_{3-4}$. Since for high-$T_c$ materials the Hall angle $\alpha_\phi$ is very small [17], we neglect its influence and suppose hereafter that $V_\phi \parallel \boldsymbol{J} \times \boldsymbol{B}$. A similar proportionality to $\Delta L_{3-4}$ holds for the quasiparticle component of the voltage, namely $W_q \propto \alpha_q \rho_n \Delta L_{3-4}$, measured on this bar sample ($\rho_n$ is the normal state resistivity). Notice thus that in a bar sample geometry, $W_q$ and $W_\phi$ have the same polarity with respect to the reversed current direction and both are insensitive to the change of the field direction. This implies some difficulties in an attempt to deconvolute the quasiparticle and fluxon core induced components to the total voltage $W$ measured across the contact pair of a bar sample in the mixed state regime, especially when $\alpha_\phi$ is very small.

At the same time, in a disk sample geometry (see insert in Fig. 2), the circular motion of the flux lines is set in by the current density, $\boldsymbol{J}$, and magnetic field, $\boldsymbol{B}$, dependent Lorentz force, $\boldsymbol{F}_L \propto \boldsymbol{J} \times \boldsymbol{B}$, which acts against various viscous mechanisms impeding the process [1]. This geometry has been successfully applied to studying the Hall effect in type I superconductors [1, 18]. When the transport current (for fixed $T$ and $B$) passes radially between the aim of the disk sample and its center, the resulting voltage drop reads $W = W_q + W_\phi = \alpha_q W_n(r) + W_\phi(r)$. With the radial arrangements of the contact pairs, it can be easily verified that $W_n(r)$ and $W_\phi(r)$ follow markedly distinct functional forms of the radial coordinate dependence, namely $W_n(r) = G(J, T) D(r)$ and $W_\phi(r) = g(J, T, B) d(r)$ with $d(r) \neq D(r)$. Observe that with one voltage contact fixed at $R_{out}$ (see insert in Fig. 2), the signal $W$ depends on the position of the radial coordinate $r$ for the second contact of the pair. For the case of $W_n$, assuming a uniform normal state resistivity and perfect equipotentiality of the leads attached to the rim, a standard text-book [13] suggests that

$$W_n(r) = \frac{1}{R_{\text{out}}} \int_0^{2\pi} d\phi \int_r^{R_{\text{out}}} r' \, dr' \, E_n(r'), \qquad (2)$$

where

$$E_n(r' > r) = \rho_n R_{\text{out}} \int_{-\pi}^{\pi} \sin\theta \, d\theta \frac{J(r')}{|\mathbf{r} - \mathbf{r}'|} = 2\rho_n R_{\text{out}} \frac{J(r')}{r'}. \qquad (3)$$

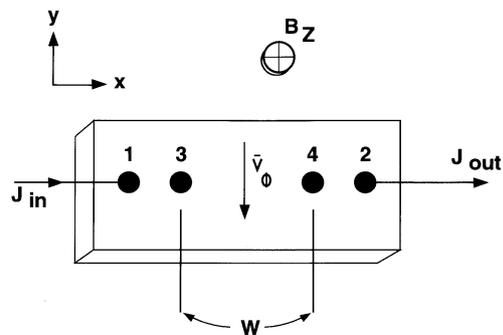

**Fig. 1.** A bar sample geometry; current density $J_{in}$ is injected via contact 1 and removed via contact 2 along the X-direction on a wide planar face. A magnetic field B is applied perpendicularly to the plane thus along the Z-direction. The voltage drop, $W = W_q + W_\phi$ is detected between the contact pair 3–4. The direction of the vortex velocity, $V_\phi$, is along the Y-axis, assuming a vanishing Hall angle, $\alpha_\phi$

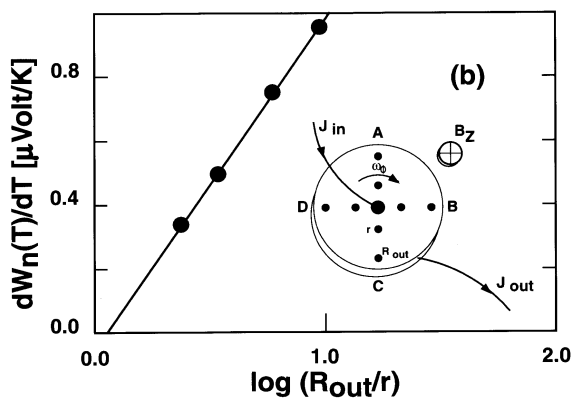

**Fig. 2.** A plot of $dW_n(T)/dT$ versus $\log(R_{out}/r)$ demonstrates the quality of equipotentiality along the disk perimeter for the normal state regime and the ($R_{out}/r$)-dependence. Insert (b): The disk sample geometry. The electrical current flows between the center of the disk and the rim. Voltage drops are measured through electrodes placed radially at various distances of the disk center and with different distances apart. The radial component of the current density $J(\sim 1/r)$ sets in a circular motion (shown by $\omega_\phi$) of the vortex system when the external magnetic field, $B$, is perpendicular to the plane of the disk

In turn, the driving current density $J(r)$ is supposed to decrease with $r$ as $J(r) = A_n/r$ and is related to the net driving current $I$ as follows

$$I = \int_0^{2\pi} d\phi \int_r^{R_{\text{out}}} r \, dr \, J(r), \qquad (4)$$

finally leading to

$$J(r) = \frac{I}{2\pi R_{\text{out}} r}, \qquad (5)$$

which in view of (2) and (3) should result in the following $r$-dependence of the quasiparticle-induced potential drop

$$W_n(r) = W_{n0} \log(R_{\text{out}}/r), \qquad (6)$$

where $W_{n0} = R_n I$ with $R_n = \rho_n/R_{\text{out}}$. This relation is remarkably verified in Fig. 2a. This means that $\rho_n$ is linear in $T$.

On the other hand, the flux-core motion induced contribution $W_\phi$ reads

$$W_\phi(r) = \frac{1}{R_{\text{out}}} \int_0^{2\pi} d\phi \int_r^{R_{\text{out}}} r' \, dr' \, E_\phi(r'), \qquad (7)$$

where

$$\mathbf{E}_\phi(r) = -\mathbf{V}_\phi(r) \times \boldsymbol{\phi}_0 n_\phi(r) \qquad (8)$$

is the electric field induced by the vortex flow with $\boldsymbol{\phi}_0$ being a vector parallel to the vortex lines the modulus of which is the flux quantum $\phi_0$, and $n_\phi(r)$ is the vortex flux density.

Assuming that the vortex transport is due to a diffusion process in which vortices move from the center of the disk to its perimeter, we arrive at the following diffusion equation for the vortex density $n_\phi(r)$

$$n_\phi(r) \mathbf{V}_\phi(r) = -D \frac{\partial n_\phi}{\partial \mathbf{r}}, \qquad (9)$$

where $D$ is a vortex diffusion coefficient.

Since the radial current density $J(r)$ decreases as $1/r$ (see (5)), the local vortex velocity is expected to follow the same dependence, that is $|V_\phi(r)| = D/r$, which in view of (9) results in a linear radial dependence of the local vortex density, $n_\phi(r) = A_\phi r$. Taking into account the normalization condition

$$N_\phi = \int_0^{2\pi} d\phi \int_r^{R_{\text{out}}} r \, dr \, n_\phi(r), \qquad (10)$$

where $N_\phi = BS/\phi_0$ is the net number of vortices inside a disk at a given value of the applied magnetic field $B$, and $S = \pi R_{\text{out}}^2$, the vortex density reads

$$n_\phi(r) = \left(\frac{3B}{2\phi_0}\right)\left(\frac{r}{R_{\text{out}}}\right). \qquad (11)$$

In view of (7) and (8), this results in the following $r$-dependence of the vortex-induced potential drop

$$W_\phi(r) = \frac{3\pi^2 DB}{2S}(R_{\text{out}}^2 - r^2) \qquad (12)$$

Notice that the result suggested by the above equation can be obtained also using a phase slip concept as introduced by Anderson[2, 12]. Indeed, when $n$ vortices (per time unit) cross the line connecting a pair of potentiometric leads, a phase difference between two contacts appears as a voltage drop [2] $W_\phi = \phi_0 n$. For the Corbino disk configuration (Fig. 2b) and under a steady state condition, characterized by an effective crossing rate, $1/\tau_\phi = 3\pi D/2S$, we get $n = N_\phi \Delta S/S\tau_\phi$, where $\Delta S = \pi(R_{\text{out}}^2 - r^2)$. Thus, the vortex-induced voltage drop for such a sample geometry can be presented as

$$W_\phi(r) = \Delta SB/\tau_\phi \equiv \gamma_\phi \Delta S \qquad (13)$$

Given $d(r) \neq D(r)$, we can deconvolute the relative strengths of the $W_q$ and $W_\phi$ terms when the disk sample geometry is thus applied. We should emphasize that $W_\phi$ results from the circular motion of vortices in the direction perpendicular to the line connecting the $r - R_{\text{out}}$ points for each pair. We note briefly that the requirement for the radial current flow is satisfied when the resistivity of the current contacts (shown as $J_{\text{in}}$ and $J_{\text{out}}$ in Fig. 1) is smaller in the normal state than that of the material the disk is made of. Since the normal state resistivity of most of the HTSs is higher than, say that of gold, we have evaporated a 70 nm gold layer at the rim of the disk and in its center. A somewhat similar condition (which also has been quite satisfactorily fulfilled in our case) applies to the superconducting state as well. It is important to mention that $W_\phi$ measured on the disk sample will have no contribution whatsoever from any "quasiparticle" related dissipation mechanisms, provided that all the contact leads are (ideally) on radial lines of the disk.

We now describe briefly the experimental procedure involved in this study. The sample geometry and contact arrangements are given in the insert of Fig. 2 and described in the figure caption. The dimensions of the disk sample used were: diameter $2R = 1.1$ cm and thickness $d = 0.2$ cm. Silver paste was used to attach 36 AWG copper wires to the previously evaporated golden pads ($\approx 0.5$ mm$^2$) for the current and potentiometric contacts on a polycrystalline melt-textured YBCO sample. The voltage signal was detected with the accuracy of 10 nV. The temperature of the sample was read and controlled by calibrated Lake Shore Carbon Glass resistors and maintained constant inside a sealed vacuum can using He exchange gas environment in a Superconducting Janis Cryostat. The mid-point of the zero-field resistance curves was detected at $T_c(0) = (87.5 \pm 0.1)$K with the zero-field transition width of $(2.5 \pm 0.1)$K. We believe that such a relatively low critical temperature and broad transition width reflect an essential deviation from (nominal) oxygen stoichiometry in our sample and suggest the possibility of the local depression of the order parameter near the oxygen depleted regions (Cf. Ref. 19). A dc magnetic field was used to excite the $W_\phi$ voltage component for variety of transport current densities, $J$. For the area of the central electrode of the order of 1 mm × 1 mm, the maximum current density in the sample (corresponding to the driving current $I = 100$ mA) was $10^8$ A/m², which is much less than $J_c(0) = 10^8$ A/m², the zero-field and zero-temperature critical current density for this material. Possible thermally induced offset voltages were eliminated by flipping the polarity of $J$.

## III. Results and discussion

The following main steps for the data analysis were used. First, for each pair of radial contacts, we find $W_n(T)$ below $T_c$. This is done by polynomial fits to the potentiometric drops detected for the sample when the temperature is well above $T_c$. Second, with the measured $W$ and the just found $W_n$, we use (1) to write:

$$W/W_n = \alpha_q + W_\phi/W_n \tag{14}$$

for each radial pair of contacts. Since the radial dependence of $W_\phi$ (see (13)) is different from that of $W_n$, we plot (at fixed $T$, $B$, and $I$) the left hand side of (14) as a function of $(R_{out}^2 - r^2)/W_n$ for each pair of contacts shown in Fig. 2. Typical data for $B = 8.29$ T and $B = 0.59$ T (and for $I = 100$ mA radial current) are presented in Fig. 3. The relation (13) is thus remarkably verified. In this figure we demonstrate the result of such a construction (see (14)) for several fixed temperatures and observe that the intercepts of $W/W_n$ axis by the straight line through the data points produce $\alpha_q$ which, as expected, is an increasing function of temperature. The slope of each line produces $\gamma_\phi = W_\phi/\Delta S$, see (13). The full-range temperature behavior of $\gamma_\phi$ and $\alpha_q$ data are shown in Fig. 4a, b.

### A. Quasiparticles

Let us first briefly discuss the temperature behavior of the relative (to $W_n$) quasiparticle contribution, $\alpha_q(T, B)$.

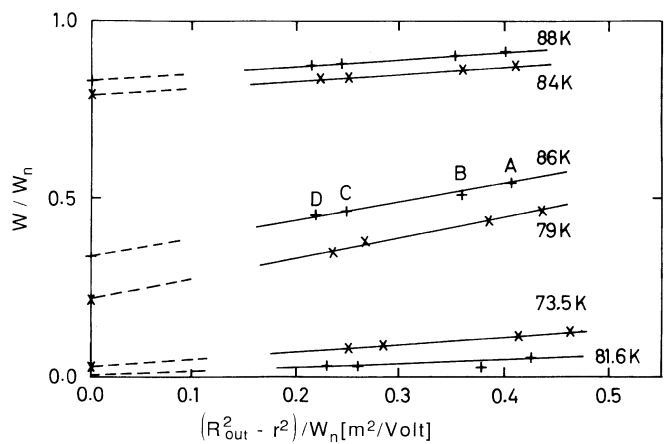

**Fig. 3.** Numerical method for extracting $\gamma_\phi$ and $\alpha_q$ (see (14)) using, respectively, the slope and the intercept with $W/W_n$ of the extrapolation of the straight lines. The $W/W_n$ data as a function of $(R_{out}^2 - r^2)/W_n$ for each pair of contacts are shown for $I = 0.1$ A and $B = 8.29$ T(×), and $B = 0.59$ T(+) for three temperatures around 80 K and 86 K of the respective data sets

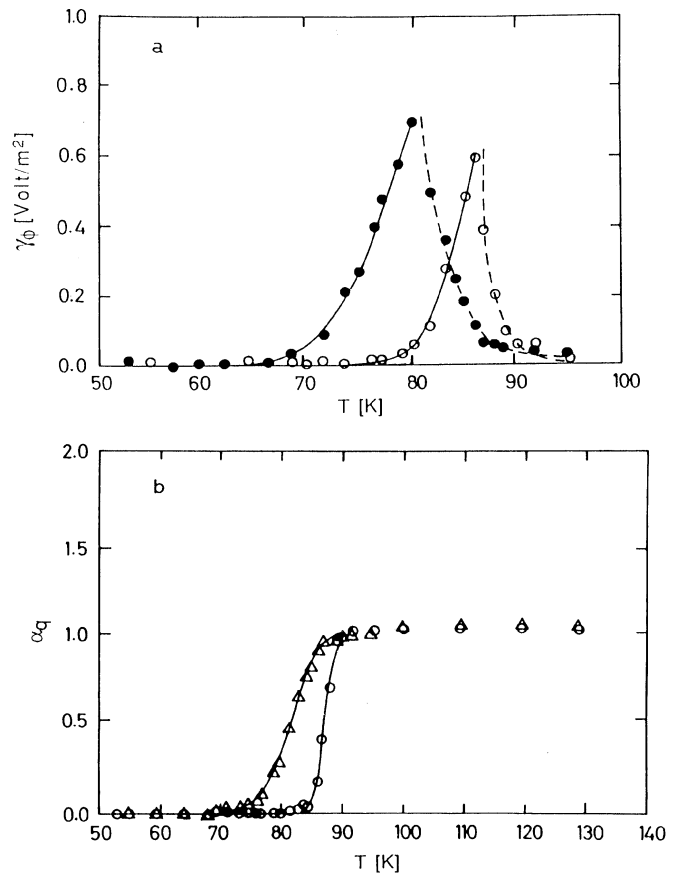

**Fig. 4.** Temperature dependence of **a** $\gamma_\phi$ and **b** $\alpha_q$ extracted by the method of Fig. 3 for $I = 0.1$ A with **a** $B = 8.29$ T(●) and $B = 0.59$ T(○); **b** $B = 0$ T(○) and $B = 8.29$ T(△). The *solid lines* through the data points for $\alpha_q(T, B)$ are the best fits using the proximity-mediated expression (see (15)). The *solid lines* through the data points for $\gamma_\phi(T, B)$ (below $T_c(B)$) are the best fits using the TAFF model expression (see (16) and (23)). The *dashed lines* through the data points for $\gamma_\phi(T, B)$ (above $T_c(B)$) are the best fits based on the Aslamazov-Larkin paraconductivity expression (see (24))

According to the BCS theory [2, 3], we could expect that just below $T_c(B)$, $\alpha_q(T, B) \approx 1 - \Delta(T, B)/2k_B T$, where $\Delta(T, B) \approx \Delta_0 \sqrt{1 - T/T_c(B)}$ is the gap parameter near $T_c(B)$. In contrast to this prediction, Fig. 4b suggests an exponential (rather than power-like) freezing out of the quasiparticle contribution. To understand such an unusual behavior of $\alpha_q(T, B)$, we come to the conclusion that it may be strongly affected by the weak-link structure in the sample, as first discussed in [20]. Here this is most likely due to an intragrain lack of oxygen stoichiometry (since $T_c$ is less than 90 K) and thus to some intrinsic (twin structure) weak-links. If this conjecture is right, we can expect that the quasiparticle contribution we measure in our experiments is related to the fluctuation-assisted resistive dissipation in the broken coherence phase near such superconductor-nonsuperconductor-superconductor (SNS)-like based junctions (see, e.g., a rather comprehensive review on the Lorentz force independent dissipation mechanisms in HTS by Kadowaki and co-workers [10] and further references therein), namely:

$$\alpha_q(T, B) = \{I_0[E_w(T, B)/2k_B T]\}^{-2}, \tag{15}$$

where $E_w(T, B) = E_w(0, B)(1 - t^2)^n$ is the Josephson ($n = 1$) or proximity ($n = 2$) energy of such a SNS contact, $I_0(x)$ is the normalized Bessel function, and $t = T/T_c(B)$. Figure 4b shows the fit (solid lines) of $\alpha_q(T, B)$ according to the above equation. It can be easily verified that the best fit suggests a proximity-like (rather than Josephson-like) coupling with $n = 2$ and the field-dependent quasiparticle activation energy of the form: $E_w(0, B) = E_{w0} \exp(-B/B_0)$, with $B_{w0} = (0.06 \pm 0.01)$ eV and $B_0 = (7.9 \pm 0.1)$ T (the so-called proximity breakdown field, see below). Notice that all these figures are in reasonable agreement with the known data for YBCO [10, 19].

### B. Vortices

Turning to the discussion of the flux induced voltage drop, $W_\phi$, we notice that since the $W_\phi$ contribution is essentially due to the vortex motion, in what follows we can use the so-called thermally assisted flux flow (TAFF) expression [21] for low transport current densities (such that $J \ll J_c(0)$) to analyze the field and temperature dependence of $\gamma_\phi$ below $T_c(B)$ [the fluctuation region, above $T_c(B)$, will be considered below]. Hence, the dissipative potential drop due to the flux motion can be presented in the form [21]

$$\gamma_\phi(J, T, B) = \gamma_0(J, B) \exp[-\beta U(T, B)], \quad (16)$$

where

$$\gamma_0(J, B) = [\beta_c U(T, B)][J/J_c(T)](\hbar B/mS)(w/r_p). \quad (17)$$

Here, $U(T, B)$ is an effective activation energy, $w$ is the hopping distance of the vortex line, $r_p$ is the range of the pinning potential, $J_c(T)$ is the zero-field critical current density, $m$ is the free electron mass, $\beta \equiv 1/k_B T$, and $\beta_c \equiv 1/k_B T_c(B)$. In turn, the activation energy can be cast into the form [21]:

$$U(T, B) = f(T, B) V_c r_p, \quad (18)$$

where $V_c$ is the coherence volume of a flux bundle, and $f(T, B) = J_c(T, B)B$ is the pinning force density.

In the so-called "amorphous" limit of the collective pinning, the coherence volume can be approximated by the expression [22]: $V_c \approx a^2(B)L_c$ with $a(B) \approx \sqrt{\phi_0/B}$ being the vortex lattice spacing, and $L_c$ the correlation length the flux line. For isolated vortex cores [21, 22], when $B < B_{c2}$, the range of pinning potential is of the order of the superconducting coherence length, i.e. $r_p \approx \xi(T)$, and the hopping distance (i.e., the distance by which the flux bundle of volume $V_c$ hops in a single thermally activated jump) is $w \approx a(B)$.

Based on the earlier suggested picture of intrinsic weak links (much activated by the non-uniform driving currents), we can expect that the temperature and field behavior of the critical currents in our sample will be dominated by the atomic scale proximity effects. On the other hand, due to an inevitable increase of the vortex viscosity coefficient near the NS boundary, the very existence of these intrinsic SNS contacts will impede the rigid circular motion of the flux lines and, in turn, result in a proximity controlled activation energy (pinning potential) for TAFF regime. Notice that the effects associated with the presence of SNS like intrinsic contacts are well documented for both conventional [23, 24] and high-$T_c$ superconductors [19, 25],

We recall that the proximity induced critical current density in SNS type structure is known [23, 24] to obey the following relation (in the low field limit, when $B < 1/\beta e D_N$):

$$J_c(T, B) = J_c(T) \exp(-B/B_0), \quad (19)$$

where

$$J_c(T) = J_c(0)(1 - t^2)^2. \quad (20)$$

Here, $B_0 = \hbar v_F/2\sqrt{3} e d_N D_N$ is the so called proximity breakdown field, with $D_N$, $v_F$, and $d_N$ being the diffusion length, Fermi velocity, and the thickness of the normal region of the SNS contact, respectively; $t = T/T_c(B)$.

In view of (18)–(20), and taking into account that $\xi(T) = \xi_0/\sqrt{1-t^2}$, the proximity affected activation energy reads:

$$U(T, B) = U(0, B)(1 - t^2)^{3/2}, \quad (21)$$

where

$$U(0, B) = U_0 \exp(-B/B_0) \quad (22)$$

and $U_0 = J_c(0)\phi_0 L_c \xi_0$.

Turning now to (17) and (18), for the prefactor of the TAFF expression (see (16)) we obtain finally:

$$\gamma_0(J, B) = C(L_c) J \sqrt{B} \exp(-B/B_0) \quad (23)$$

with $C(L_c) = (\hbar L_c/2meS) \beta_c \phi_0^{3/2}$.

Examples of the analytical fits to $\gamma_\phi$ (below $T_c(B)$) are shown in Fig. 4. The best fits to all our field data for $\gamma_\phi$ produce an exponential field dependence of the activation energy (see (22)). The plot of $\log[U(0, B)]$ versus $B$, which is shown in Fig. 5, produces a straight line, with the slope of $1/B_0$, which results in the proximity breakdown field, $B_0 = (7.8 \pm 0.1)$ T, which is very close to the one we found above for the proximity affected quasiparticle contribution. It is worthwhile to mention also that similar values of $B_0$ were found in YBCO thin films [19] as well as twinned polycrystals [25] within practically the same field and temperature regimes of study. The corresponding value of $U_0$ found by our procedure is $U_0 = (0.34 \pm 0.05)$ eV, in reasonable agreement with reported data on YBCO [26]. Observed that (23) suggests that the plot of $\log[\gamma_0/\sqrt{B}]$ as a function of $B$ should essentially result in a straight line parallel to the one for the field dependence of $\log[U(0, B)]$. This is verified by the second set of data of Fig. 5, thus demonstrating the self-consistency of our data analysis with the experimental results. Using typical values of $\xi_0 \approx \xi_{ab}(0) = 1.5$ nm, $J_c(0) = 10^8$ A/m$^2$, and inserting $U_0 = 0.3$ eV, (22) brings about a reasonable estimate of the correlation length along the vortex line, $L_c \approx 200$ μm. We can also estimate the approximate thickness of the normal region, $d_N$, of the intrinsic SNS contact. Indeed, with $D_N = l v_F/3$, $B_0 = 7.8$ T, and taking the typical value of the electron mean free path at $T_c$ to be [27] $l \approx 15$ nm, we find in terms of (19) that $d_N \approx 5$ nm,

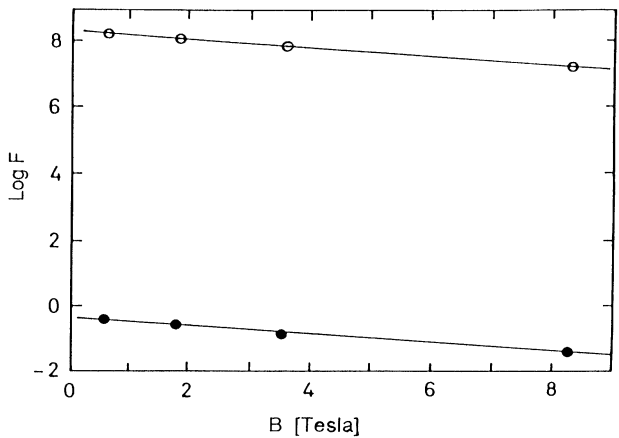

**Fig. 5.** The lines labeled (○) and (●) are, respectively, for the field dependent activation energy, $U(0, B)$, [with $F = U(0, B)$ in Kelvin] and the prefactor $\gamma_0$ of (23) [with $F = \gamma_0/B^{1/2}$, where $\gamma_0$ and $B$ are normalized to there respective units, Volt/m$^2$ and Tesla]. Observe that the best linear fits, which are indicated by the straight lines through the data points produce, as evident from (22), the same slope $1/B_0$, with $B_0 = (7.8 \pm 0.1)$ T

which is of the order of the characteristic size of oxygen-depleted regions in YBCO [25], i.e. of twin spacing. We have also verified that for the current densities such that $J \ll J_c(0)$, $\gamma_0$ is a linear function of the driving current for $I = 10, 30,$ and $100$ mA, in accord with the prediction of (23).

### C. Paraconductivity above the critical temperature

Finally, to complete our analysis, let us briefly discuss the region above $T_c(B)$. According to the conventional theory of paraconductivity [2], the fluctuation contribution to the potential drop registered there can be represented as $W^{fl} = W - W_n$ or, in view of (14), $W^{fl}/W_n = W_\phi/W_n - (1 - \alpha_q)$. Since, however, $\alpha_q \approx 1$ above $T_c(B)$ (see Fig. 4b), this region appears to be dominated by the macroscopic fluctuations (see also [28]) accompanying the vortex core motion, i.e. $W^{fl} \approx \gamma_\phi^{fl} \Delta S$. For two-dimensional Gaussian fluctuations which are expected for our sample geometry, Aslamazov-Larkin theory predicts [2]:

$$\gamma_\phi^{fl}(T, B) = \gamma_0(J, B) \frac{T}{T - T_c(B)}, \quad (24)$$

where $\gamma_0(J, B)$ is still given by (23) but with $C = C(d)$ [where $d$ is the sample thickness]. The dashed lines through the data points in Fig. 4b show the forced fit for $\gamma_\phi$ above $T_c(B)$ according to (24), with $d = (2.3 \pm 0.5)$ mm, which is very close to the actual thickness of the disk used in our experiments.

### IV. Conclusion

In conclusion, to deconvolute the quasiparticle and vortex related contributions to the mixed state dissipation of melt-textured YBCO, a disk sample geometry has been used. The analysis of our experimental results was based on the assumption of markedly different radial dependencies of the voltages in the total potentiometric signal detected. When the thermally activated weak-link and flux flow (TAFF) models were used to analyze our experimental results, the corresponding activation energies for the quasiparticle excitations and flux creep processes were found to exhibit an exponential field dependence, consistent with proximity affected dissipation mechanism. The intrinsic nature of the latter phenomenon is argued to be due to the extensive weak links structure presumably present in the sample and activated by the non-uniform current density profile generated in the disk geometry. The paraconductivity region appears to be dominated by the macroscopic fluctuations accompanying the vortex core motion.

Technical assistance and contribution made by P. de Villiers to setting and performing the experimental part of this study is greatly appreciated. We wish to thank P.J. Van Der Schyff and E. Rood for their excellent technical support in due course of investigation. MA thanks J.R. Clem for his interest and correspondence. Part of this work has been financially supported through the Impulse Program on High Temperature Superconductors of Belgium Federal Services for Scientific, Technological and Cultural (SSTC) Affairs under the Contract No. SU/02/013. Support of the Ministery of High Education and Scientific Research through ARC (94-99/174) grant is acknowledged.